\documentclass[aps,prl,preprint,superscriptaddress]{revtex4}
\usepackage{bm}
\begin{document}
\title{Emergent gauge field from self-referencing phase factor on many-body wave functions and superconductivity}
\author{Hiroyasu Koizumi}
\affiliation{Division of Quantum Condensed Matter Physics, Center for Computational Sciences, University of Tsukuba, Tsukuba, Ibaraki 305-8577, Japan}
\date{\today}
\begin{abstract}
Superconductivity is a phenomenon where electrical current flows without friction. The current standard theory for it is the BCS (Bardeen-Cooper-Schrieffer) theory, which explains it as due to the energy gap formation by the electron-pairing, and the key ingredient for the supercurrent generation is the gauge symmetry breaking brought about by it. 
It was thought that superconductivity was fully understood by this standard theory; however, the discovery of superconductivity in cuprates in 1986 changed this situation, showing a number of experimental results that contradict the standard theory. 
It is also notable that the standard theory contradicts in the supecurrent carrier mass in the London moment; the predicted mass is an effective mass, while the experimental value is the free electron mass.
The above contradictions suggest the necessity for a fundamental revision for the theory of superconductivity.

 Here we show that the required revision may be achieved by using the Berry phase formalism.
 It was developed after the establishment of the BCS theory, and provides a way to detect emergent gauge fields. A self-referencing phase factor on the wave function detected by the Berry phase formalism 
 explains the supercurrent generation in the conventional and cuprate superconductors.
 It gives rise to a gauge field that enables the gauge symmetry breaking in the standard theory interpretation.
\end{abstract}
\maketitle
The standard theory of superconductivity is based on the BCS theory. According to this theory,
the origin of superconductivity is the energy gap formation due to electron-pairing \cite{BCS1957}. The rigidity of the wave function against external perturbations envisaged by London \cite{London1950} is realized by this energy gap, and one of the hallmarks of superconductivity, the exclusion of a
magnetic field from the superconductor (the Meissner effect), is explained by this rigidity.

The BCS used the following variational state vector, 
\begin{eqnarray}
|{\rm BCS}\rangle =
\prod_{\bf k}(u_{\bf k}+v_{\bf k}c^{\dagger}_{{\bf k} \uparrow} c^{\dagger}_{-{\bf k} \downarrow} )|{\rm vac} \rangle
\label{theta}
\end{eqnarray}
to take into account the electron pairing effect,
 where $|{\rm vac} \rangle$ is the vacuum state, $c^{\dagger}_{{\bf k} \sigma}$ is the creation operator for the conduction electron of effective mass $m^{\ast}$ with the wave vector ${\bf k}$ and spin $\sigma$, and $u_{\bf k}$ and  $v_{\bf k}$ are variational parameters. The obtained energy gap provides the rigidity of the superconducting state against external perturbations, explains energy gap related experimental results, and providing a method to calculate the superconducting transition temperatures $T_c$ as the energy gap formation temperature. 

A salient feature of the BCS state vector in Eq.~(\ref{theta}) is that it dose not satisfy the conservation of the particle number; this means that the number of electrons in an isolated superconductor is not fixed.
This non-conservation property is a crucial ingredient of the standard superconductivity formalism; it breaks the global $U(1)$ gauge symmetry, and this gauge symmetry breaking is actually the key to explain the Meissner effect and supercurrent flow in superconductors \cite{Anderson1958a,Anderson1958b,Nambu1960}. 

In spite of the success of such a particle number non-conserving formalism,  it is sensible to think that the particle number is conserved in an isolated superconductor, thus, the usage of the particle number non-conservaing state vector appears at odd.
In fact, some researchers have been speculated that the use of this particle number non-conserving formalism will be eventually replaced by a formalism that conserves the particle number \cite{Feynman,LeggettBook}.

In 1986, high $T_c$ superconductivity was discovered in cuprates \cite{Muller1986}. It shows a number of contradictions against the standard theory. There are many experiments that indicate electron-pairing is relevant; however, it appears that the electron pair formation occurs at temperatures much higher than $T_c$ \cite{Zhou2019}, and $T_c$ does not correspond to the pairing energy gap formation temperature \cite{Kivelson95}. 

Another important point of the cuprate superconductivity is that the normal state above $T_c$ is very anomalous; despite the strong correlation between electrons \cite{AndersonBook} and small lattice polaron formation \cite{Bianconi}, the carrier mass obtained from the London moment measurement is the free electron mass \cite{Verheijen1990}, and any evidence of the usual effects of phonon or impurity scattering is not seen in the relevant regions of the phase diagram \cite{Anderson2000}.
It is also worth mentioning that Planckian timescale dissipation \cite{Zaanen2004} is deduced from the Uemura's law (the proportionality between the superfluid density and $T_c$) \cite{Uemura1989}, Homes' law (the proportionality between the superfluid density and the product of $T_c$ and electric conductivity at $T=T_c$) \cite{Homes2004}, and Tanner's law (the proportionality between the superfluid density and the doped hole density in the low doping samples) \cite{Tanner1998}, resulting the $T$ linear electric resistivity up to very high temperature, which is quadratic to $T$ in ordinary metals \cite{AndersonBook}. 

Moreover, the inelastic neutron scattering experiment reveals the coexistence of  superconductivity and magnetism \cite{Xu2009}, although the magnetism is known to be harmful for the conventional superconductors. The observation of loop currents at temperatures much higher than $T_c$ \cite{Kerr1,Nernst} is also a very peculiar fact since such loop currents are usually associated with superconductivity that occurs at temperatures below $T_c$.

Besides the cuprate superconductivity, actually, the standard theory has been known to contradict in the mass of the supercurrent carrier, although this discrepancy is not widely appreciated; the standard theory predicts the mass appears in the London moment is the effective mass $m^{\ast}$, however, the experimental results always indicate it to be the free electron mass $m_e$; for example, the free electron mass is obtained for the conventional superconductors \cite{Hildebrandt1964,Zimmerman1965,Brickman1969,Tate1989,Tate1990}, the high $T_{\rm c}$ cuprates \cite{VERHEIJEN1990a,Verheijen1990}, and
heavy fermion superconductors \cite{Sanzari1996}.

Recently, this London-moment-mass contradiction is lifted using a formalism that uses the Berry connection from many-body wave functions \cite{koizumi2019,koizumi2021}. In this theory, the many-body wave function is given by
\begin{eqnarray}
\Psi({\bf x}_1, \cdots, {\bf x}_N)=\exp \left(  i \sum_{j=1}^{N} \int_0^{{\bf r}_j} {\bf A}^{\rm MB}_{\Psi}({\bf r}') \cdot d{\bf r}'  \right)
\Psi_0({\bf x}_1, \cdots, {\bf x}_N)
\label{single-valued}
\end{eqnarray}
where ${\bf x}_i$ collectively denotes the coordinate ${\bf r}_i$ and the spin $\sigma_i$ of the $i$th electron, $\Psi_0({\bf x}_1, \cdots, {\bf x}_N)$ is the currentless wave function that is obtained by the energy minimization, and the factor $\exp \left(  i \sum_{j=1}^{N} \int_0^{{\bf r}_j} {\bf A}^{\rm MB}_{\Psi}({\bf r}') \cdot d{\bf r}'  \right)$ is the one arising from the Berry connection ${\bf A}^{\rm MB}_{\Psi}$, which is defined using $\Psi({\bf x}_1, \cdots, {\bf x}_N)$ by 
 \begin{eqnarray}
{\bf A}^{\rm MB}_{\Psi}({\bf r})&=&-i  \int d\sigma_1 d{\bf x}_{2} \cdots d{\bf x}_{N}
{ {\Psi^{\ast}({\bf r}, \sigma_1, {\bf x}_{2}, \cdots, {\bf x}_{N})} \over {|C_{\Psi}({\bf r})|^{{1 \over 2}}}}
 \nabla_{\bf r}
{ {\Psi({\bf r}, \sigma_1, {\bf x}_{2}, \cdots, {\bf x}_{N})} \over {|C_{\Psi}({\bf r})|^{{1 \over 2}}}}
\nonumber
\\
&=&{1 \over {\hbar\rho({\bf r})}}{\rm Re} \left\{ 
 \int d\sigma_1  \cdots d{\bf x}_{N}
 \Psi^{\ast}({\bf r}, \sigma_1, \cdots, {\bf x}_{N})
  {\bf p}_{\bf r}
\Psi({\bf r}, \sigma_1, \cdots, {\bf x}_{N}) \right\}
\nonumber
\\
\label{Afic}
\end{eqnarray}
with $|C_{\Psi}({\bf r})|$ being the normalization constant
 \begin{eqnarray}
|C_{\Psi}({\bf r})|=\int d\sigma_1 d{\bf x}_{2} \cdots d{\bf x}_{N}\Psi({\bf r},\sigma_1, {\bf x}_{2}, \cdots)\Psi^{\ast}({\bf r}, \sigma_1, {\bf x}_{2}, \cdots)
\end{eqnarray}
and $\rho$ being the electron density obtained from $\Psi$.

A peculiar point is that the factor $e^{- i \sum_j\int_{0}^{{\bf r}_j} {\bf A}_{\Psi}^{\rm MB}({\bf r}',t) \cdot d{\bf r}'}$ in $\Psi$ is a self-referencing quantity obtained from the wave function $\Psi$ itself.
Its effect cannot be included by the perturbation theory or the configuration-interaction method. A method to obtain it is developed by the present author using the single-valuedness condition of $\Psi({\bf x}_1, \cdots, {\bf x}_N)$ as a function of ${\bf r}_1, \cdots, {\bf r}_N$, and the local charge conservation condition \cite{koizumi2020c,Koizumi2021c}. 

Although the Berry phase (or Berry connection) was derived originally in the context of the adiabatic approximation \cite{Berry}, it
can be calculated just from the wave function without the adiabatic assumption; it can be used 
as a tool to detect emergent gauge fields. 
In other words, the gauge field given by the vector potential, ${\bf A}^{\rm MB}_{\Psi}$ is detected using the
Berry phase formalism. 

Since $\Psi_0$ is a currentless state, the factor $e^{- i \sum_j\int_{0}^{{\bf r}_j} {\bf A}_{\Psi}^{\rm MB}({\bf r}',t) \cdot d{\bf r}'}$  adds  the following kinetic energy term
 \begin{eqnarray}
\int { \hbar^2 \over {2 m_e}}({\bf A}^{\rm MB}_{\Psi})^2 \rho({\bf r}) d^3 r
\label{Supercurrent-E}
\end{eqnarray}
to the energy calculated by $\Psi_0$ \cite{Bohm1949}.
 Supercurrent is generated by this term. 
 Usually, the state described by $\Psi$ is higher in energy than the currentless state $\Psi_0$,
 as is known as the ``Bloch's theorem''  \cite{Bohm1949}. 
To realize superconductivity, some how his situation is overcome.
In the following, we show that a way out of this situation is to consider the second quantization. 

Let us define an angular variable $\chi$  
given by
\begin{eqnarray}
{ {\chi({\bf r})} \over 2}= \int^{{\bf r}}_0 {\bf A}_{\Psi}^{\rm MB}({\bf r}') \cdot d{\bf r}' 
\end{eqnarray}

The canonical conjugate momentum of $\chi$, $\pi_{\chi}$, is obtained as
\begin{eqnarray}
\pi_{\chi}=-{ \hbar \over 2} \rho
\end{eqnarray}

The canonical quantization condition for $\chi$ and $\rho$ is given by
\begin{eqnarray}
[\chi({\bf r}),\pi_{\chi} ({\bf r}')]=i\hbar \delta({\bf r}-{\bf r}')
\end{eqnarray}

Then, we can construct the following boson field operators 
\begin{eqnarray}
\psi_{\chi}^{\dagger}({\bf r})=\sqrt{\rho({\bf r})}e^{-{ i \over 2}
\chi({\bf r})}, \quad \psi_{\chi}({\bf r})=e^{{ i \over 2}\chi({\bf r})}\sqrt{\rho({\bf r})}
\end{eqnarray}
that satisfy the following commutation relation
\begin{eqnarray}
[ \psi_{\chi}({\bf r}), \psi_{\chi}^{\dagger}({\bf r}')]=\delta({\bf r}-{\bf r}')
\end{eqnarray}

Byy integrating them over the space,
\begin{eqnarray}
B^{\dagger}_{\chi} =\int d^3r  \psi_{\chi}^{\dagger}({\bf r}),
\quad B_{\chi} =\int d^3r \psi_{\chi}({\bf r})
\end{eqnarray}
we obtain boson operators $B^{\dagger}_{\chi}$ and $B_{\chi}$
that satisfy
\begin{eqnarray}
[ B_{\chi}, B_{\chi}^{\dagger}]=1
\end{eqnarray}

Then, the following number operator 
\begin{eqnarray}
\hat{N}_{\chi}=B_{\chi}^{\dagger} B_{\chi}
\end{eqnarray}
is obtained. We identify $\chi$ as the variable for the collective mode arising from the non-trivial Berry connection, and $\rho$ as the number density of electrons participating in this mode, and $\hat{N}_{\chi}$ as
the number operator of the electrons participating in this mode.

We construct a phase operator $\hat{X}$ that is conjugate to $\hat{N}_{\chi}$
 through the relations
 \begin{eqnarray}
B_{\chi}^{\dagger} =\sqrt{\hat{N}_{\chi}}e^{-i\hat{X}}, \quad B_{\chi} =e^{i\hat{X}}
\sqrt{\hat{N}_{\chi}}
\end{eqnarray}

The phase and number operators have the following commutation relation
\begin{eqnarray}
[e^{i\hat{X}}, \hat{N}_{\chi}] =e^{i\hat{X}}
\end{eqnarray}
indicating $e^{ i\hat{X}}$ is the number changing operator that decreases the
number of electrons participating in the collective mode by two.

Using $e^{ i\hat{X}}$, a state vector similar to the one in Eq.~(\ref{theta}) is constructed 
\begin{eqnarray}
|{\rm Gnd}\rangle =
\prod_{\bf k}(u_{\bf k}+v_{\bf k} c^{\dagger}_{{\bf k} \uparrow} c^{\dagger}_{-{\bf k} \downarrow}e^{i \hat{X}} )|{\rm Cnd} \rangle
\label{chi}
\end{eqnarray}
 where the state vector $|{\rm Cnd} \rangle$ corresponds to the state given by the wave function $\Psi$.
The operator
$c^{\dagger}_{{\bf k} \uparrow} c^{\dagger}_{-{\bf k} \downarrow}e^{i \hat{X}}$
 acting on $|{\rm Cnd} \rangle$ annihilates two electrons participating in the collective mode, and creates electrons in single-particle states.
  
By employing the approximation used by BCS \cite{BCS1957}, we obtain
\begin{eqnarray}
u_{\bf k}^2={ 1 \over 2} \left(1 +{{{E}_{\bf k}} \over {\sqrt{ E_{\bf k}^2+ \Delta^2 }}} \right), \quad
v_{\bf k}^2={ 1 \over 2} \left(1 -{{{E}_{\bf k}} \over {\sqrt{ E_{\bf k }^2+ \Delta^2}}} \right), \quad
\Delta\approx 2 \hbar \omega_c e^{-{ 1 \over {N(0)V}}}
\end{eqnarray}
from the state vector $|{\rm Gnd}\rangle$, 
where $N(0)$ is the density of states at the Fermi energy, $V$ is the parameter for the attractive interaction between electrons around the Fermi energy, and $\hbar \omega_c$ is the cut-off energy for this attractive interaction; ${E}_{\bf k}$ is the single-particle energy measured from the Fermi level, and the pairing energy gap is given by
\begin{eqnarray}
\Delta=V \sum_{\bf k} \langle {\rm Gnd}|e^{-i \hat{X}}c_{-{\bf k} \downarrow} c_{{\bf k} \uparrow} | {\rm Gnd } \rangle =V \sum_{\bf k} u_{\bf k} v_{\bf k}
\label{Delta}
\end{eqnarray}

Actually, the above corresponds to 
\begin{eqnarray}
\Delta=V \sum_{\bf k} \langle {\rm BCS}|c_{-{\bf k} \downarrow} c_{{\bf k} \uparrow} | {\rm BCS}\rangle =V\sum_{\bf k} u_{\bf k} v_{\bf k} 
\end{eqnarray}
in the BCS theory \cite{BCS1957}. In this theory, the effect of $e^{-i \hat{X}}$ in Eq.~(\ref{Delta}) is included using the particle number non-conserving property of $|{\rm BCS}\rangle$.

The total energy in the present theory is a sum of of the kinetic energy from $e^{- i \sum_j\int_{0}^{{\bf r}_j} {\bf A}_{\Psi}^{\rm MB}({\bf r}',t) \cdot d{\bf r}'}$ and the total energy from $|{\rm Gnd} \rangle$,
\begin{eqnarray}
E_{\rm tot}&=&\int d^3 r {{\hbar^2 \rho({\bf r})} \over {2 m_e}}
\left({ 1 \over 2}\nabla \chi
\right)^2+2\sum_{\bf k } {E}_{\bf k} v_{\bf k}^2-{{\Delta^2} \over V}
\nonumber
\\
&\approx&{{N(0)\Delta} \over {\cal V}} \int  d^3r {{\hbar^2 \left(\nabla \chi
\right)^2 } \over {4 m_e}}
-{1 \over 2}N(0)V\Delta^2
\end{eqnarray}
where $\nabla \chi$ and $\rho$ are expectation values of the corresponding operators for the ground state; we neglected the coordinate dependence $\rho$, and the number of electrons in the collective mode is calculated as
\begin{eqnarray}
\int  d^3r  \rho({\bf r})&=&{\cal V} \rho
\nonumber
\\
 &=&\sum_{ {E}_{\bf k }\leq 0}u^2_{\bf k}
\nonumber
\\
&=&
2N(0)\int_{-\hbar \omega_c}^0 
{ 1 \over 2} \left(1 +{{x} \over {\sqrt{ x^2+ \Delta^2}}} \right)dx
\nonumber
\\
&=&N(0)\left(\hbar \omega_c +\sqrt{\Delta^2}-\sqrt{ \hbar^2 \omega_c^2+ \Delta^2} \right)
\nonumber
\\
&\approx& N(0) \Delta
\label{nonzero}
\end{eqnarray}
where ${\cal V}$ is the volume of the system.
If the energy gap formation makes the energy of the current carrying state lower than that of the currentless state, the superconducting state will be realized. 

When a magnetic field ${\bf B}^{\rm em}$ exists, the vector potential from magnetic field ${\bf A}^{\rm em}$ (${\bf B}^{\rm em}=\nabla \times {\bf A}^{\rm em}$) appears in addition to the gauge field ${\bf A}^{\rm MB}_{\Psi}$. Then, the kinetic energy of the collective mode becomes
\begin{eqnarray}
E_{\chi}=\int d^3 r {{\hbar^2 \rho({\bf r})} \over {2 m_e}}
\left({ 1 \over 2}\nabla \chi+{e \over \hbar}{\bf A}^{\rm em}
\right)^2
\label{Echi}
\end{eqnarray}

Then, the supercurrent density is given by
\begin{eqnarray}
{\bf j}&=&-{{\partial E_{\chi}} \over {\partial {\bf A}^{\rm em}}}
\nonumber
\\
&=&-{{e^2 \rho({\bf r})} \over {m_e}}
\left({ \hbar \over {2e}}\nabla \chi+{\bf A}^{\rm em}
\right)
\end{eqnarray}
This is a diamagnetic current that explains the Meissner effect. The presence of the angular variable $\chi$ with period $2\pi$ and characterized by a topological integer (the {\em winding number}) 
\begin{eqnarray}
w_C[\chi]={1 \over {2\pi}}\oint_C \nabla \chi \cdot d{\bf r}
\end{eqnarray}
where $C$ is a closed path, 
explains the flux quantum ${ h \over {2e}}$. 

The velocity field associated with the above supercurrent is $
{\bf v}_s
={{e} \over {m_e}}
\left({ \hbar \over {2e}}\nabla \chi+{\bf A}^{\rm em}
\right)
$. 
The mass in this formula, the free electron mass $m_e$, explain lifts the London-moment-mass discrepancy \cite{koizumi2021}. Moreover, the state vector in Eq.~(\ref{chi}) conserves the particle number.

The angular variable $\chi$ in the conventional superconductors is argue to arise from spin-twisting itinerant motion of electrons that circulates around a section of the Fermi surface \cite{koizumi2020}; thereby, $e^{-i \hat{X}}$ in Eq.~(\ref{Delta}) is realized. The pairing gap $\Delta$ formation occurs simultaneously with the
appearance of $e^{-i \hat{X}}$ and the stabilization of $\nabla \chi$, thus, $\Delta$ can be used as the order parameter of the superconducting phase.

Now, we consider the cuprate superconductivity. 
The parent compound is an insulator, called the ``Mott insulator'' where the strong electron-electron repulsion makes it an antiferromagnetic insulator.
Upon hole doping, the electric conduction appears.
The optical spectral measurement exhibits 4:1 ratio of mid-infrared to free-carrier spectral weight.
The origin of the mid-infrared spectral weight is known to be the small polaron formation of the doped holes \cite{Bianconi,Miyaki2008}. 

A theory predicts spin-vortex formation in the bulk around the
small polarons formed by the hole doping \cite{Koizumi2011,HKoizumi2013}.
The existence of spin-vortices explains the hourglass-shaped magnetic excitation spectrum \cite{Neutron,Berciu2004b,Hidekata2011}. Note that the small polaron formation is suppressed in the surface region, probably due to the absence of the charge reservoir layer. In the surface region, the $d$-wave electron pairing is realized.

In the cuprate, $\Psi_0$ part of $\Psi$ in Eq.~(\ref{single-valued}) is the wave function for this spin-vortex equipped states. It is a multi-valued function with respect to electron coordinates. Then, the factor $\exp \left(  i \sum_{j=1}^{N} \int_0^{{\bf r}_j} {\bf A}^{\rm MB}_{\Psi}({\bf r}') \cdot d{\bf r}'  \right)$ arises in $\Psi$ 
to satisfy the single-valued condition.

As a consequence, loop currents are generated by it around the small polarons.  These loop currents are called the {\em  spin-vortex-induced loop currents (SVILCs)} \cite{koizumi2020c,Koizumi2021c}; they are characterized by the topological integer, the {\em winding number}.  
The existence of loop currents are well established from the polar Kerr effect measurement \cite{Kerr1}, enhanced Nernst effect measurement \cite{Nernst}, and the neutron scattering measurement \cite{neutron2015}. 
  
In the situation where SVILCs are generated in the bulk of the cuprate, $\nabla \chi$ is given by
\begin{eqnarray}
\nabla \chi=\sum_{i=1}^{N_{hole}} \nabla \chi_{h_i}
\end{eqnarray}
where $ \nabla \chi_{h_i}$ is the contribution from the $i$th hole, and $N_{hole}$ is the number of holes;
each $\nabla \chi_{h_i}$ is characterized by a winding number around its singularity (or the vortex flux).

Then, $E_{\chi}$ in Eq.~(\ref{Echi}) for the two-dimensional CuO$_2$ layer, omitting ${\bf A}^{\rm em}$ is given by \begin{eqnarray}
E_{\chi}&=&\int d^2 r {{\hbar^2 \rho({\bf r})} \over {2 m_e}}
\left({ 1 \over 2} \sum_{i=1}^{N_{hole}} \nabla \chi_{h_i}
\right)^2
\nonumber
\\
&\approx&
\int d^2 r {{\hbar^2 \rho({\bf r})} \over {8 m_e}}
\left[ \sum_{i=1}^{N_{hole}} (\nabla \chi_{h_i})^2 +2\sum_{\langle i, j \rangle } (\nabla \chi_{h_i}) \cdot (\nabla \chi_{h_j})
\right]
\end{eqnarray}
where the sum over $\langle i, j \rangle$ means the sum over nearest neighbor pairs.

We assume that the small polarons with doped holes at their cores form a two-dimensional square lattice, approximately. We denote that the contribution from $\nabla \chi_{h_i}$ on one of the sub-lattices by ${\bf S}_{h_i}$ and that from $\nabla \chi_{h_j}$ on the other by $-{\bf S}_{h_j}$,
and assume that the energy is approximately given by
\begin{eqnarray}
E_{\chi}
&\approx&
 A \sum_{i=1}^{N_{hole}} ({\bf S}_{h_i})^2-B\sum_{\langle i, j \rangle } {\bf S}_{h_i} \cdot 
 {\bf S}_{h_j}
\end{eqnarray}

We treat ${\bf S}_{h_i}$ as the spin lying in the $xy$ plane, and express the nearest neighbor interaction between them by ${\bf S}_{h_i}\cdot {\bf S}_{h_j}={\bf S}^2 \cos( \phi_i -\phi_j)$.
As a consequence, 
\begin{eqnarray}
E_{\chi}
&\approx&
 A \sum_{i=1}^{N_{hole}} ({\bf S}_{h_i})^2-B\sum_{\langle i, j \rangle } {\bf S}^2 \cos( \phi_i -\phi_j)
 \nonumber
 \\
 &\approx&
{\rm const.}
 +{ {J n_{hole}} \over 2} \int d^2r  (\nabla \phi)^2
 \label{Echi2}
\end{eqnarray}
is obtained, 
where the fact is used that the total energy depends on the number of nearest neighbor site pairs in a unit area, thus, is proportional to $n_{hole}$.

A transition similar to the Berezinskii-Kosterlitz-Thouless type transition \cite{Berezinskii,KST} is expected for the above model \cite{HKoizumi2015B,Koizumi2017}; thus, the transition
temperature will satisfy
\begin{eqnarray}
T_c \propto  n_{hole}
\end{eqnarray}
 This explains the ``Uemura's law'' \cite{Uemura1989}.

The supercurrent density is given by 
\begin{eqnarray}
{\bf j}=-{{e \hbar \rho({\bf r})} \over {2 m_e}}\sum_{i=1}^{N_{hole}}\nabla \chi_{h_i}
=-{{e \hbar \rho({\bf r}) N_{hole}} \over {2 m_e}} \overline{\nabla \chi},
\quad
\overline{\nabla \chi}=
{1 \over {N_{hole}}}\sum_{i=1}^{N_{hole}}\nabla \chi_{h_i}
\label{current-density}
\end{eqnarray}
If $\overline{\nabla \chi}$ is almost constant by changing the hole density in a corse-grained cell where ${\bf j}$ is evaluated, the supercurrent will be proportional to $n_{hole}$.
Thus, the proportionality of the superfluid density 
\begin{eqnarray}
n_s \propto n_{hole}
\label{propto}
\end{eqnarray}
 is realized. This explains the ``Tanner's law'' \cite{Tanner1998}.

The electric conductivity is given by 
\begin{eqnarray}
 \sigma_{\nu x}(\omega)=\lim_{\epsilon \rightarrow +0}
 \int_0^{\infty}e^{i \omega t -\epsilon t}dt\int_0^{(k_BT)^{-1}}\langle j_x(-i \hbar \lambda) j_\nu (t) \rangle_{\rm eq}d\lambda
\end{eqnarray}
according to the linear response theory \cite{Kubo1957}, 
where $t$ dependent current density component $j_\nu (t)$ is defined as
\begin{eqnarray}
j_{\nu}(t)=e^{i H_0  t/ \hbar } j_{\nu}e^{-i H_0 t/\hbar }
\end{eqnarray}
with $H_0$ being the non-perturbed Hamiltonian,
and $\langle \hat{O} \rangle_{\rm eq}$ being the equilibrium value of the operator $\hat{O}$.

Using the current density in Eq.~(\ref{current-density}), it is written as
\begin{eqnarray}
 \sigma_{xx}(0)=
 \int_0^{\infty} dt { 1 \over {k_B T}}\langle j_x(0) j_x(t) \rangle_{\rm eq}
 \label{sigma}
\end{eqnarray}
where the ``classical approximation'' $\hbar \lambda \approx 0$ is employed; this approximation is known to be valid in the critical region of the phase diagram \cite{Sachdev}.

Evaluating this at $T=T_c$, we obtain
\begin{eqnarray}
 k_B T_c\sigma(T_c)=
 \int_0^{\infty} dt \langle j_x(0) j_x(t) \rangle_{\rm eq}
 \label{Home}
\end{eqnarray}
where $\sigma(T_c)$ is $\sigma_{xx}(0)$ at $T=T_c$.
Note that the time-dependence of $j_x$ contains the abrupt change of $\partial_x \chi_{h_i}(t)$ due to the change of the winding numbers of $\chi_{h_i}$.

At $T=T_c$, the critical fluctuation of $\partial_x \chi_{h_i}(t)$ will give rise to the random fluctuation (or a $\delta$-function type time-correlation),
\begin{eqnarray}
 \langle j_x(0) j_x(t) \rangle_{\rm eq}&=& {{ \hbar^2 e^2 \rho^2} \over { 4m_e^2}} \sum_{i} \langle \partial_x \chi_{h_i}(0) \partial_x \chi_{h_i}(t)
\rangle_{\rm eq}
\nonumber
\\
&=& C n_{hole} \delta(t)
\end{eqnarray}
where $C$ is a constant.

Thus, Eq.~(\ref{Home}) yields
\begin{eqnarray}
 k_B T_c\sigma(T_c)=C n_{hole} 
 \label{Home2}
\end{eqnarray}
This formula explains the ``Home's law''  \cite{Homes2004} since $n_{hole}$ is proportional to $n_s$ in Eq.~(\ref{propto}).
The random fluctuation is expected to persist above $T_c$ since the transition is a Berezinskii-Kosterlitz-Thouless like. Then, Eq.~(\ref{sigma}) explains the $T$-linear resistivity observed \cite{AndersonBook}. 

According to the present theory, the superconducting state occurs when the gauge field given by $\nabla \chi$ is stabilized. It appears in a self-referencing phase factor on the many-body wave function. The effect of this emergent gauge field cannot be taken into account by the perturbation theory or the configuration interaction calculation customarily used in the electronic structure calculation. The effect of it is taken into account by using the particle number non-conserving formalism in the standard theory.

 Self-reference phenomena and their peculiar consequences are known in nature. The G\"{o}del's incompleteness theorem \cite{Goedel} is a famous example, and there is a speculation that ``mind'' might arise from some kind of self-referencing \cite{GEB}. The present work indicates that superconductivity is one of them.
 

\end{document}